\documentclass[conference]{IEEEtran}
\IEEEoverridecommandlockouts
\usepackage{cite}
\usepackage{amsmath,amssymb,amsfonts}
\usepackage{algorithmic}
\usepackage{graphicx}
\usepackage{textcomp}
\usepackage{xcolor}
\usepackage{multirow}
\usepackage{array}

\usepackage[numbers,sort]{natbib}
\usepackage[colorlinks = true,
            linkcolor = blue,
            urlcolor  = blue,
            citecolor = blue,
            anchorcolor = blue]{hyperref}

\def\BibTeX{{\rm B\kern-.05em{\sc i\kern-.025em b}\kern-.08em
    T\kern-.1667em\lower.7ex\hbox{E}\kern-.125emX}}
\begin{document}

\title{Teddy: Automatic Recommendation of Pythonic Idiom Usage For Pull-Based Software Projects}

\author{\IEEEauthorblockN{Purit Phan-udom\textsuperscript{\textasteriskcentered}, Naruedon Wattanakul\textsuperscript{\textasteriskcentered}, Tattiya Sakulniwat\textsuperscript{\textasteriskcentered},\\ Chaiyong Ragkhitwetsagul\textsuperscript{\textasteriskcentered}, Thanwadee Sunetnanta\textsuperscript{\textasteriskcentered}, Morakot Choetkiertikul\textsuperscript{\textasteriskcentered}, Raula Gaikovina Kula\textsuperscript{\textdagger}
}
\IEEEauthorblockA{\textit{\textsuperscript{\textasteriskcentered}Faculty of Information and Communication Technology (ICT), Mahidol University} \\
\textit{\textsuperscript{\textdagger}Nara Institute of Science and Technology (NAIST)} \\
\textit{\{purit.pha, naruedonw, tattiya.sakul\}@gmail.com, \{chaiyong.rag, thanwadee.sun, morakot.cho\}@mahidol.ac.th,} \\ \textit{raula-k@is.naist.jp}
}
}

\maketitle

\begin{abstract}
Pythonic code is idiomatic code that follows guiding principles and practices within the Python community.
Offering performance and readability benefits, Pythonic code is claimed to be widely adopted by experienced Python developers, but can be a learning curve to novice programmers. 
To aid with Pythonic learning, we create an automated tool, called Teddy, that can help checking the Pythonic idiom usage.
The tool offers a \textit{prevention mode} with Just-In-Time analysis to recommend the use of Pythonic idiom during code review and a \textit{detection mode} with historical analysis to run a thorough scan of idiomatic and non-idiomatic code.
In this paper, we first describe our tool and an evaluation of its performance.
Furthermore, we present a case study that demonstrates how to use Teddy in a real-life scenario on an Open Source project.
An evaluation shows that Teddy has high precision for detecting Pythonic idiom and non-Pythonic code. 
Using interactive visualizations, we demonstrate how novice programmers can navigate and identify Pythonic idiom and non-Pythonic code in their projects. 
Our video demo with the full interactive visualizations is available at \url{https://youtu.be/vOCQReSvBxA}.
\end{abstract}

\section{Introduction} 
According to the TIOBE index\footnote{\url{https://www.tiobe.com/tiobe-index/}} and the GitHub 2019 annual report\footnote{\url{https://octoverse.github.com/}}, Python is in the top three of the most popular programming languages.
Python emphases being elegant and easy to read\footnote{ Zen of Python at \url{https://www.python.org/dev/peps/pep-0020/}}. Similar to any other programming language, Python code also has an idiomatic \textit{``Pythonic''} way of writing code to solve a particular problem.  


In the age of social coding, such as Open Source GitHub projects, programmers find themselves in situations where they would start contributing to a project that involves Python programming. Learning and using Pythonic idioms can be a challenging task for novices who join the project~\cite{alexandru2018}. 
There exists literature that provides examples on Pythonic idioms \cite{book:Writing_Idiomatic_Python}, however, most of the learning seems to be from the communities through experience and usage examples available on forums such as Stack Overflow~\cite{alexandru2018}.
Compared to non-Pythonic code, Pythonic idioms are well-accepted form of Python programming ~\cite{web:python_3.8.0} due to many benefits.
For example, Sakulniwat et al.~\cite{Sakulniwat2019} showed that adopting the \texttt{with open} Pythonic idiom gives benefits not only in terms of readability, but also prevents memory leaks by automatically closing the file after finishing the task.

In this paper, we present \textit{Teddy}, an automated tool for recommending Pythonic idiom usage for pull-based development software projects. 
Based on two common types of code contributions \cite{Hattori2008ASE}, 
Teddy offers two modes to recommend opportunities for novices to make their code more idiomatic.
The first is \textbf{\textit{prevention mode}} with Just-In-Time (JIT) analysis.
In this mode, Teddy searches for development activities that are related to incorporation of new features and implementation (i.e., forward engineering).
We identify non-Pythonic usage during the code review process by analyzing any submitted patches in pull request's commits.
The tool then recommends a Pythonic idiom counterpart of doing the same task to the developers.  
The second is \textbf{\textit{detection mode}} with historical analysis.
Teddy analyses the whole commit history of a given project and looks for both occurrences of Pythonic idiom (Py) and non-Pythonic (NPy) code snippets during software maintenance (i.e., re-engineering, corrective and management).

Teddy produces an interactive visualization to show the usage of Pythonic idioms over time based on the detected Py and NPy code in all the commits.  
To demonstrate our tool, we present a case study to demonstrate Teddy in a real-life scenario.
An evaluation shows that Teddy has high precision for detecting idiomatic and non-idiomatic Python code. 
Using visualizations, we demonstrate how novice programmers navigate and identify Py or NPy code.



\section{Architectural Design}
\begin{figure}[htbp]
    \centering
    \includegraphics[width=0.9\columnwidth]{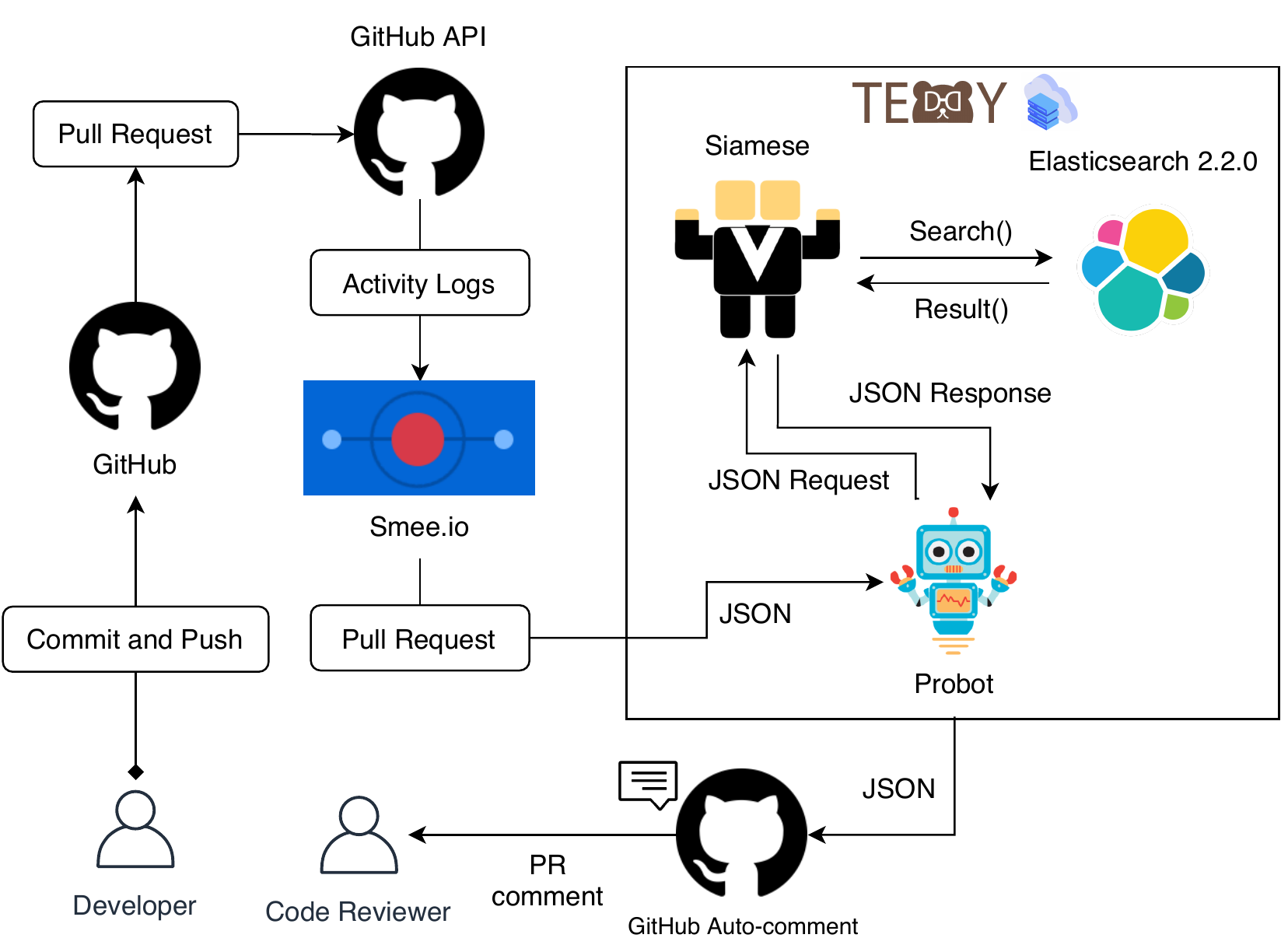}
    \caption{Architecture of the prevention mode (JIT analysis)}
    \label{fig:system-architect-prevent}
\end{figure}

\begin{figure}[htbp]
    \centering
    \includegraphics[width=0.9\columnwidth]{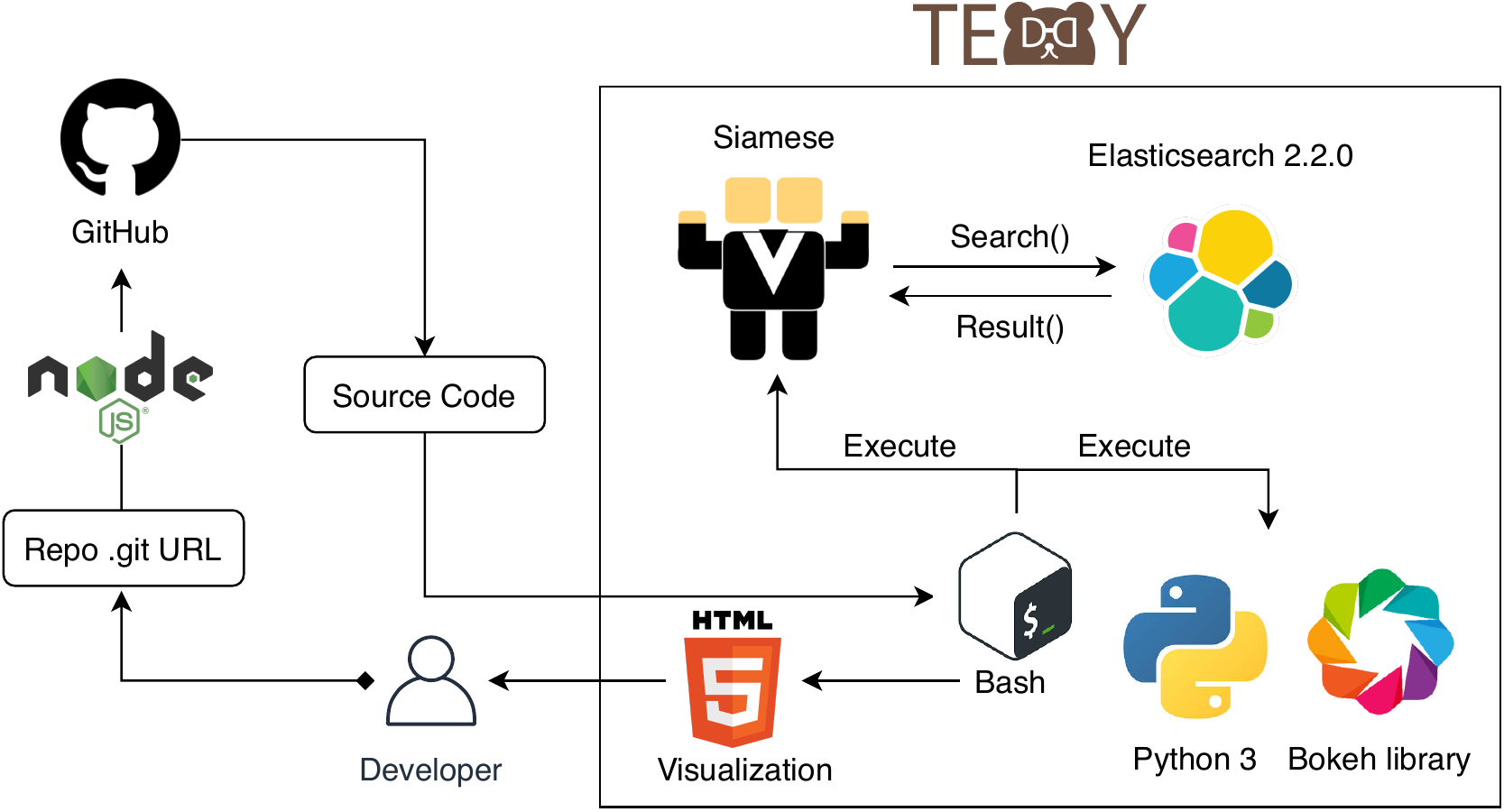}
    \caption{Architecture of the detection mode (historical analysis)}
    \label{fig:system-architect-detect}
\end{figure}

Teddy's architecture has two  aspects: Just-In-Time Pythonic idiom analysis (prevention mode) as shown in Fig.~\ref{fig:system-architect-prevent} and Pythonic idiom evolving in the history (detection mode) in Fig. \ref{fig:system-architect-detect}. 
The prevention mode aims to give early feedback to code reviewers by identifying Py and NPy code snippets from GitHub pull requests and the results are delivered as pull request comments. The detection mode provides the Pythonic idiom visualization from the past commits which software practitioners can oversee the Pythonic adoption. 

Our design is divided into three main components: the Pythonic idiom and non-Pythonic code database, the idiom matching, and the idiom usage interactive visualization.



\textit{1. The Py/NPy database}.
Table~\ref{table:ips-nips} shows four examples of Py and NPy cases taken from our built database.  
In total, our database contains 113 code snippets of which 55 are Pys and 58 are NPys.
The complete list of our labelled types, with their actual code snippets is available from our website at \url{https://muict-seru.github.io/icsme20-teddy-tooldemo}.

Our database is based on Python programming community's best practices  \cite{book:Writing_Idiomatic_Python,web:github_pythonic}.
We identified and labelled ten different types of Pythonic idioms to be included in our database.
The ten types are based on their functionalities and include \textit{dictionary comprehension, enumerate, file reading statement, list comprehension, if statement, string formatting, set, tuple, variable swapping}, and \textit{code formatting}, where each of them has both Py and NPy code snippets in our database.
We thus have 20 original code snippets (10 Pys and 10 NPys). We then modify these original code snippets to create a variety of patterns, i.e., data augmentation, in each type, such as renaming identifiers and changing data types.

\begin{table}[]
    \centering
    \caption{Examples of the Pythonic idiom (Py) and Non-pythonic (NPy) code studied in this project}
    \label{table:ips-nips}
    \resizebox{.5\textwidth}{!}{ %
    \begin{tabular}{|c|p{2cm}|p{6cm}|}
    \hline
    Type & Name & Description \\ \hline
    Py & enumerate & for-loop iteration using \texttt{enumerate} function \\ \hline
    Py & file reading statement & Using \texttt{with open() as ...} to open a file \\ \hline
    NPy & variable swapping & Using a temporary variable to swap two variables' values\\ \hline
    NPy & code formatting & Using ';' to put more than one statement in a \newline single line\\ \hline
    \end{tabular} %
    }
\end{table}

\textit{2. Idiom matching based on clone search}.
The Py/NPy database serves as a corpus in the idiom matching. We employ a code clone search technique, called Siamese~\cite{Ragkhitwetsagul2019}, to determine whether code snippets from a software project similar to Py or NPy code snippets in our database. Siamese allows flexible matching at different levels of code abstraction using its multiple code representation technique. Thus, it offers a capability of detecting Pys and NPys with several modifications (i.e., Type-2 and Type-3 clones).


\textit{3. Idiom usage interactive visualization}.
To manage large software projects, we employ an interactive visualization to show the identified instances of Pythonic idiom and Non-Pythonic usage.
Our tool allows a user to pan and zoom over many files to inspect Py and NPy usage and their evolution over commits in a project.



\section{Implementation}
\textbf{Prevention mode}. In this mode, Teddy is implemented as an automated bot during GitHub pull requests (Figure~\ref{fig:system-architect-prevent}). The non-Pythonic code snippets in the Py/NPy database are stored in Siamese's code search index (using Elasticsearch). Then, Teddy connects with GitHub through Smee.io\footnote{https://smee.io}, a webhook service, and uses the code changes in a pull request's commits as Siamese's search queries. The queries are matched with the NPys in the code search index using the idiom matching component. A query that is a clone to any NPy is returned with a recommendation of its Py counterpart from the Py/NPy database. The overall results are formatted into a friendly message (i.e., with an explanation of a recommended Py snippet to replace the NPy one). Teddy then performs an auto-comment in the pull request by the Probot\footnote{https://github.com/probot/probot} service.


\textbf{Detection mode}. In this mode, Teddy first obtains the source code from a given GitHub URL by cloning the repository (Figure~\ref{fig:system-architect-detect}). Then, the tool iterates through the commit versions of the cloned repository, from the first to the last commit.
Siamese creates a code search index from the project source code in each commit, and injects the Py/NPy code snippets from the Py/NPy database as search queries. The results are combined and fed into the visualization component, using the Bokeh\footnote{https://github.com/bokeh/bokeh} library, where a scatter-plot graph is generated.

\section{Evaluation}
\label{sec:evaluation}

\textbf{Evaluation Metrics}.
We prepared the Siamese search index with a ground-truth dataset containing Python code snippets that are labelled as Py, NPy, and normal code. 
We performed the idiom matching by sending 113 queries in our Py/NPy database to Siamese and checked how many Py/NPy code snippets were correctly retrieved. 
From the results of 113 queries, we computed mean average precision (MAP), mean reciprocal rank (MRR), query recall (QR) and overall recall (OR). 
The calculation of MAP and MRR follow the definitions in information retrieval~\cite{Manning2009}. 
For recall, we defined two types of recall, QR and OR, as follows:

\textit{Query Recall (QR)} determines how complete is the retrieved relevant items with respect to the subset of queries with \textit{non-empty results} (i.e., returned queries). 
QR for a set of returned queries $r$ is defined as: $QR = \frac{1}{|r|}\sum_{i=1}^{|r|}\frac{|R_i|}{|A_i|}$ where $R_i$ is the set of retrieved relevant items and $A_i$ the set of all relevant items for the $i$-th returned query, respectively. 

\textit{Overall Recall (OR)} evaluates how complete is the number of relevant items retrieved with respect to the entire query set. 
In contrast to QR, OR also considers the queries with empty results returned. OR for a query set $R$ is defined as $OR = \frac{1}{|R|}\sum_{i=1}^{|R|}\frac{|R_i|}{|A_i|}$ where $R_i$ is the set of retrieved relevant items and $A_i$ the set of all relevant items for the $i$-th query.

\textbf{Ground-Truth}.  We collected Python source code files, containing Py and NPy snippets inside, from external sources, mainly from three GitHub repositories of Flask, Tensorflow TFX, Ipyparallel, and programming websites\footnote{geeksforgeeks.org, programiz.com, tutorialpoint.com, journaldev.com, beginnersbook.com, stackoverflow.com, and note.nkmk.me}.
First, we searched for Py and NPy code snippets within the three GitHub repositories by using CCGrep~\cite{ccgrep}, a code clone detector that can detect clones by giving regular expressions. 
Table~\ref{table:eval-dataset} shows the three groups of Python code files. 
The first group, the normal code, are Python code snippets that do not contain Py nor NPy code snippets inside. 
The second group, the Py group, are files that contain a Pythonic idiom. 
The third group, the NPy group, are files that contain a non-Pythonic code snippet.

\begin{center}
    \begin{table}[htbp]
        \centering
        \caption{The ground-truth dataset}
        \label{table:eval-dataset}
        \begin{tabular}{|l|p{4cm}|r|} 
         \hline
         File group & Description & Files \\ \hline
         Normal & Python code without Py or NPy & 30 \\ \hline 
         Py & Pythonic idiom code snippets & 20 \\ \hline
         NPy & Non-Pythonic code snippets & 20\\ \hline
         \multicolumn{2}{|l|}{\textbf{Total}} & \textbf{70}\\
         \hline
        \end{tabular}
    \end{table}
\end{center}


    
    

\textbf{Accuracy Results}.
Table~\ref{tab:teddy-evaluation} presents four configurations $C_1$, $C_2$, $C_3$, and $C_4$ where we varied two of Siamese's parameters including the similarity measure (n-gram token ratio and fuzzywuzzy) and the similarity thresholds ($T_0$, $T_1$, $T_2$, $T_3$) of the four code representations in Siamese.
We found that different similarity measures offer different performance on the idiom matching. In general, idiom matching using n-gram token ratio gains higher MAP and MRR values comparing to those of fuzzywuzzy. The similarity threshold also plays an important role in the accuracy of idiom matching. Choosing low similarity thresholds (i.e., $C_1$ and $C_2$) resulted in high recall but low MAP, which mean a large number of false positives were retrieved. In contrast, choosing higher similarity thresholds (i.e., $C_3$, and $C_4$) gave high MAP, but we needed to sacrifice the recall. 

As a result, we found that the configuration $C_4$ gives the best idiom matching performance by offering MAP of 0.89 and MRR of 0.83.
This configuration, however, has a relatively low recall. The QR value of 0.5 shows that, for all the queries that returned results, we retrieved half of the Py and NPy code snippets. The low OR value of 0.04 came from a low number of queries that returned the results. The modified variants that we added into the Py/NPy database made them matched with only very specific patterns of Py and NPy not available in the ground-truth dataset. In the end, we prefer to get high precision to avoid presenting false positives to users.
    
\begin{table}
\centering
    \caption{Configurations of Siamese evaluated in the experiment}
    \label{tab:teddy-evaluation}
    \resizebox{.5\textwidth}{!}{ %
    \begin{tabular}{|l|c|r|r|r|r|r|r|r|r|}
    \hline
    \multirow{2}{*}{} & \multicolumn{1}{c|}{\multirow{2}{*}{Sim.}} & \multicolumn{4}{p{9em}|}{Multi-representation\newline similarity threshold} & \multicolumn{4}{c|}{Error Measures} \\ \cline{3-10} 
    & \multicolumn{1}{c|}{} & $T_0$ & $T_1$ & $T_2$ & $T_3$ & \multicolumn{1}{c|}{MAP} & \multicolumn{1}{c|}{QR} & \multicolumn{1}{c|}{OR} & \multicolumn{1}{c|}{MRR} \\ \hline
	$C_1$ & FWZ & 0 & 0 & 0 & 0 & 0.32 & 1.00 & 1.00 & 0.44 \\ \hline
	$C_2$ & NTR & 0 & 0 & 0 & 0 & 0.35 & 0.89 & 0.88 & 0.45 \\ \hline
    $C_3$ & FWZ & 40 & 40 & 40 & 40 & 0.38 & 0.47 & 0.46 & 0.41 \\ \hline
    $C_4$ & NTR & 40 & 40 & 40 & 40 & 0.89 & 0.50 & 0.04 & 0.83 \\ \hline
    \multicolumn{10}{l}{\textit{NTR = N-gram Token Ratio, FWZ = Fuzzywuzzy}}
    \end{tabular} %
	}
\end{table}

\begin{figure}
	\centering
	\includegraphics[width=\linewidth]{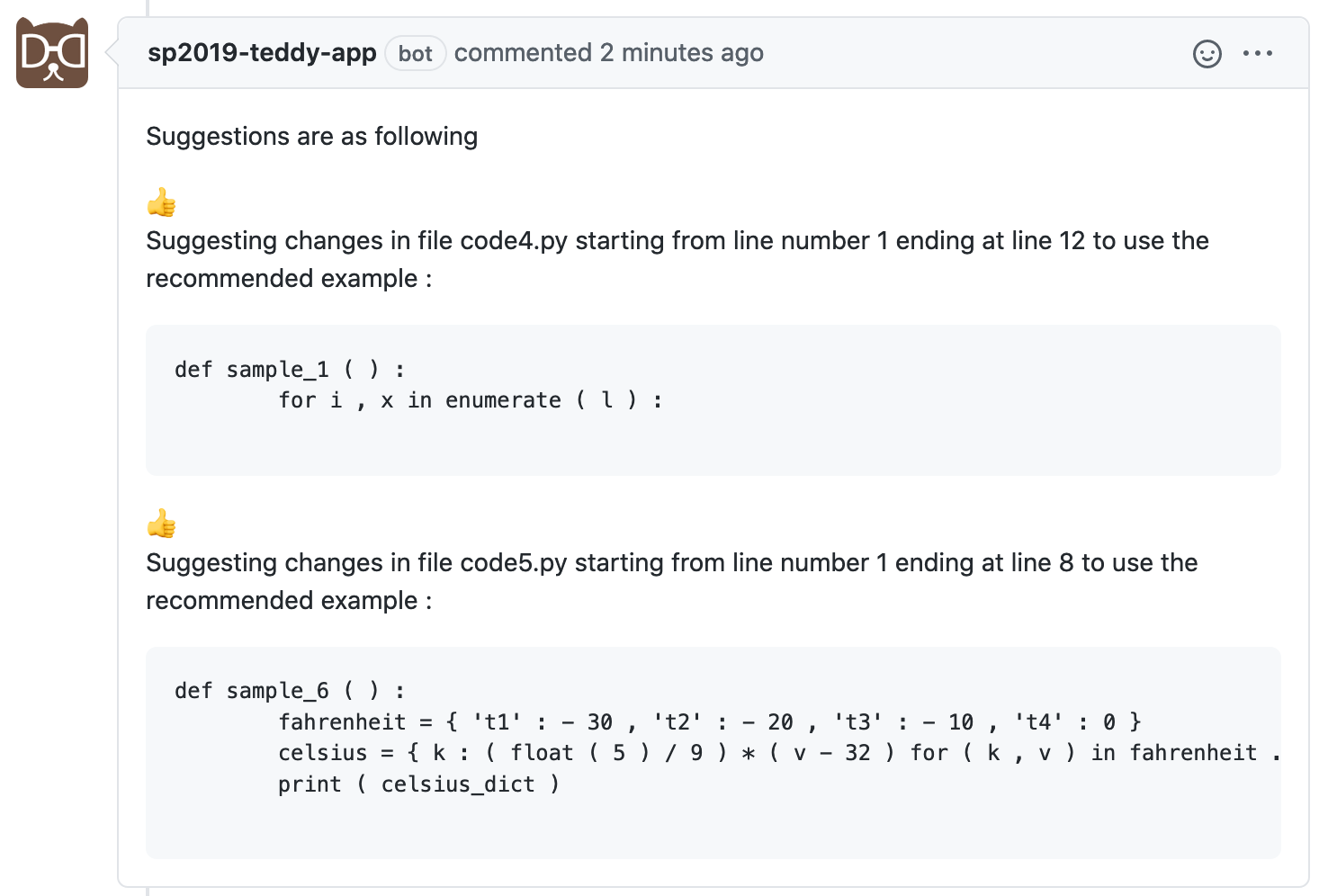}
	\caption{Result of prevention mode: Recommending Pythonic idioms in a pull request (available at~\url{https://github.com/MUICT-SERU/flask/pull/2}).}
	\label{fig:flask-pr-comment}
\end{figure}

\begin{figure*}[!]
    \centering
    \includegraphics[width=0.95\linewidth]{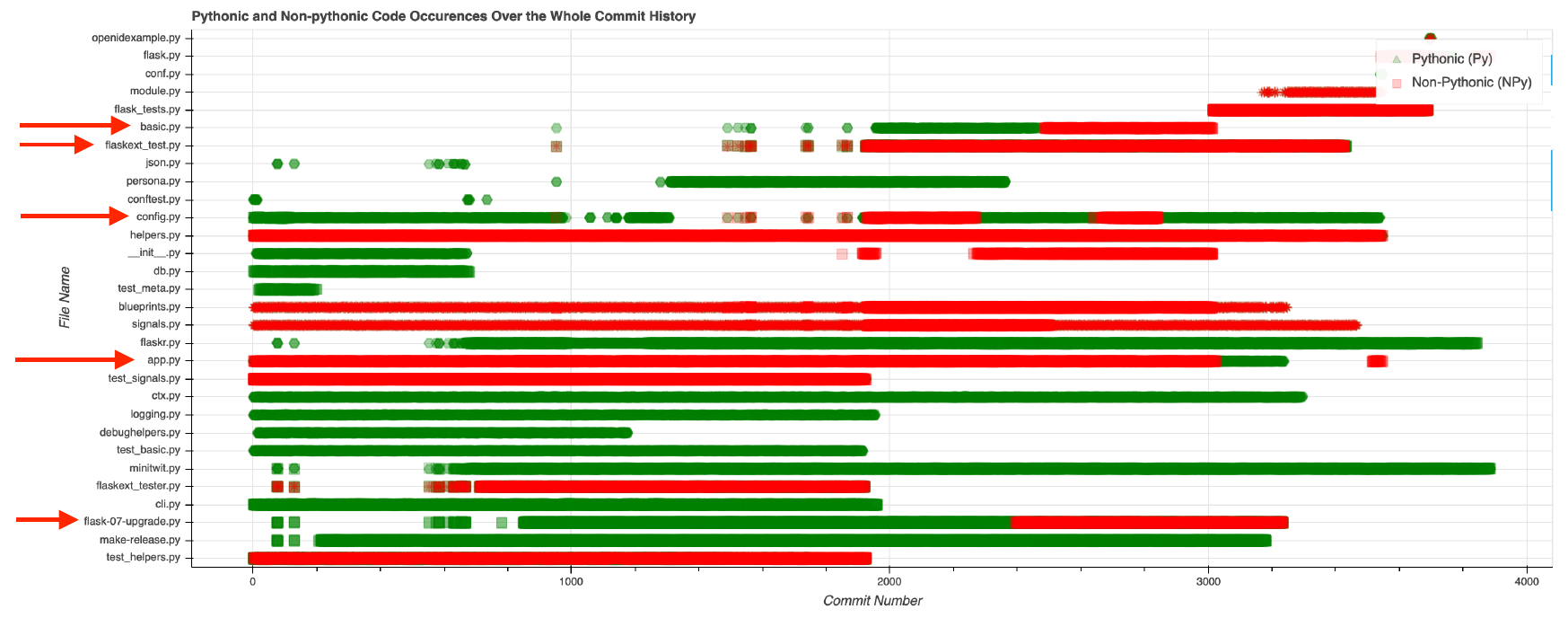}
    \caption{Result of detection mode: An interactive visualization for Py and NPy usage over the whole commit history. A full interactive sample of the visualization, which includes the pan and zoom functions, is available on the study website at \url{https://muict-seru.github.io/icsme20-teddy-tooldemo/flask.html}.}
    \label{fig:detection-flask-modified}
\end{figure*}

\textbf{Case Study on the Flask Project}.
We performed a case study on Flask, a popular web application framework written in Python, with 50.9k stars, 13.6k forks, and more than 3.9k commits on GitHub. 

In the prevention mode, we needed to simulate the situation of creating a pull request. This could not be done on the actual Flask project due to the permission to install the Teddy bot and to create a pull request. Thus, we forked the Flask project into our GitHub account first and installed the Teddy bot on the repository. Then we created a new pull request with 5 commits. Each commit contained one Python file that performs a specific task. There were two files that contained the non-Pythonic code that could be replaced with the Pythonic idiom of \textit{enumeration} and \textit{dictionary comprehension}. 
In the detection mode, we provided the GitHub repository URL of the Flask project to Teddy. Then, the Teddy tool cloned the source code of the project into its system and performed the idiom matching and visualization. 

\textbf{Teddy Outputs}.
Figure~\ref{fig:flask-pr-comment} shows that Teddy bot detected the two non-Pythonic code snippets in the pull request and gave the recommendations for their Pythonic idiom usages. 
We can see that Teddy suggested a Py example (i.e., \texttt{sample\_1} snippet) based on the code snippet in the Py/NPy database to replace the detected NPy in \texttt{code4.py}.

Figure~\ref{fig:detection-flask-modified} shows a visualization of the occurrences of Pythonic idioms and non-Pythonic code over time. 
The x-axis displays the number of commits while the y-axis displays the files containing the detected Py and NPy. Each dot represents one occurrence of Py and NPy in the project, which can be hovered on to show details of the type of Py and NPy and the line numbers (start, end) that it appears in the file. 

From our visualization, a user is able to see changes of the Pythonic idioms over time. 
For instance, the figure shows that about half of the files detected by Teddy in the project contain non-Pythonic code (red lines) that never be replaced by Pythonic idiom. In contrast, the other half of the detected files contain Pythonic idioms (green lines) that also never change to non-Pythonic code. There are a few files that show a switch from non-Pythonic code to Pythonic idioms (i.e., changing from red to green) such as \texttt{app.py} and \texttt{flaskext\_test.py}. The other direction (Py to NPy or from green to red lines) can also be observed in \texttt{basic.py} and \texttt{flask-07-upgrade.py}. Lastly, there is a file, \texttt{config.py}, that show multiple changes between Py and NPy. These files are pointed to by an arrow in the figure.

\section{Conclusion}
Teddy promotes the usage of Pythonic idioms by giving recommendations to developers.
Our evaluation shows that Teddy accurately detects Pythonic idioms and non-Pythonic code with MAP of 0.89 and MRR of 0.83. We demonstrate the tool's usefulness by a case study using Flask GitHub project.

\section*{Acknowledgement}
This work is supported by Japanese Society for the Promotion of Science (JSPS) KAKENHI Grant Numbers 18H04094 and 20K19774.

\bibliographystyle{IEEEtran}
\bibliography{references.bib} 

\end{document}